\documentclass[oribibl]{llncs}
 
\usepackage{amssymb} 
\usepackage{amsmath} 
\usepackage{url}

\usepackage{graphicx}

\makeatletter
\@twosidefalse
\makeatother

\begin{document}

\title{\textsc{SymbolicData}:\textsc{SDEval} - Benchmarking for Everyone}

\author{Albert Heinle\inst{1}, Viktor Levandovskyy\inst{2}, Andreas Nareike\inst{3}}

\institute{Cheriton School of Computer Science, University of
  Waterloo, Waterloo, Canada
\and Lehrstuhl D f\"ur Mathematik, RWTH Aachen University, Aachen, Germany
\and HTWK Leipzig, Leipzig, Germany
}

\maketitle

\begin{abstract}
  In this paper we will present \textsc{SDeval}, a software project
  that contains tools for creating and running benchmarks with a
  focus on problems in computer algebra. It is built on top of the
  \textsc{Symbolic Data} project, able to translate problems in the
  database into executable code for various computer algebra
  systems. The included tools are designed to be very flexible to use and to extend, such that they can be utilized even in contexts of other communities. With the presentation of \textsc{SDEval}, we will also address particularities of benchmarking in the field of computer algebra.

Furthermore, with \textsc{SDEval}, we provide a feasible and automatizable way of reproducing benchmarks published in current research works, which appears to be a difficult task in general due to the customizability of the available programs.

We will simultaneously present the current developments in the \textsc{Symbolic Data} project. 
\end{abstract}


\section{Introduction}

Benchmarking of software -- i.e. measuring the quality of results and the time resp. memory consumption for a given, standardized set of examples as input -- is a common way of evaluating implementations of algorithms in many areas of industry and academia. For example, common benchmarks for satisfiability modulo theorems (SMT) solvers are collected in the standard library \textsc{SMT-LIB} (\cite{barrett2010smt}), and the advantages of various solvers like \textsc{Z3} (\cite{de2008z3}) or \textsc{CVC4} (\cite{barrett2011cvc4}) are revealed with the help of those benchmarks.

Considering the field of computer algebra, there could be various
benchmarks for the different computation problems. Sometimes, one can
find common problem instances throughout papers dealing with the same
topics, but often there is no standard collection and authors use
examples best to their knowledge. For the calculation of Gr\"obner
bases for example, there is a collection of ideals that often appear
when a new or modified approach accompanied by an implementation is
presented (e.g. in \cite{sneumann2012Parallel}, the author used the
\textsc{Katsura}-$n$, $n \in \{11,12\}$ from \cite{katsura1987distribution} and
\textsc{Cyclic}-$m$, $m \in \{8,9\}$ from
\cite{bjorck2008all}). Regarding the computation on that set, the new
implementation is then compared to existing and available ones.

An outstanding systematic practice is realized by the computer algebra
lab, lead by V.~P. Gerdt, of the``Joint Institute
For Nuclear Research'', on their website about the progress in
research of computing Janet and Gr\"obner bases of complicated
polynomial systems (\url{http://invo.jinr.ru/}).

Nevertheless, in many areas there is rarely a standard test set, and often the calculated timings are hard to reconstruct due to different parameters in algorithms that can be set.

Another difficulty is the fair evaluation on how much time is
consumed; we will discuss this topic detailed in section \ref{sctn:SDEval}.

The \textsc{Symbolic Data} project (\cite{grabe2006symbolicdata}) started more than 10 years ago to collect various instances of problems, especially in computer algebra. 

In this paper, we discuss particularities concerning code quality evaluation especially for the computational algebra community and present \textsc{SDEval}, a benchmarking toolbox
written in \textsc{Python} covering the following two main tasks:
\begin{itemize}
  \item[(i)] Creating benchmark sets with the help of the problem instances provided by the \textsc{Symbolic Data} database entries.
  \item[(ii)] Running benchmarks, with a flexible (i.e. cross-community adaptable) interface that makes reproduction as simple as possible.
\end{itemize}

For item (i), we implemented for certain computational problems (e.g. calculation of a Gr\"obner basis) translators of respective problem instances in \textsc{Symbolic Data} into executable code for a set of computer algebra systems. It can be done using a graphical user interface, or alternatively via a program run in the terminal. This task addresses e.g. developers, who want to compare the running time of their implementations with those of available software without the necessity of becoming familiar with all of the available systems. Additionally, it addresses mathematicians who discovered a certain instance for a computational problem and want to examine what computer algebra systems are able to solve it and what solutions are provided, as they might differ -- dependent on the uniqueness of the result -- for the different systems.

Item (ii) has a broader range of possible uses. First of all, it
provides a way to run arbitrary programs on different
inputs. Optionally, it monitors the computations and terminates
programs automatically if they exceed a user-given time or memory
limit. Moreover, it provides the user with intermediate information on
which tasks are currently run and which have already finished, and one
can stop certain calculations manually without having to start the
whole process all over again. The information is provided through a generated \textsc{HTML} document and -- for potential automation purposes -- through an \textsc{XML} file. We chose for this part a structure that is strongly connected to a folder, which can be shared with others and after an negligible amount of adjustment to another machine, the results can be reproduced. We envision for the future that \texttt{tar}-balls of those folders would be published with computation-focused papers, so that it becomes easier to verify results of the authors.

This paper is structured as follows. Section \ref{sctn:SData} deals
with current developments regarding \textsc{Symbolic Data} and
presents a general overview of this project. The subject of Section
\ref{sctn:SDEval} will be \textsc{SDEval}. We will show how one can
use it and extend/adjust it to individual computation problems if needed. We will finish by addressing related work in Section \ref{sctn:relatedWork} and future tasks to come in Section \ref{sctn:Conclusion}.

The current version of the presented toolkit \textsc{SDEval} can be found at \url{github.com/ioah86/symbolicdata}.
The latest informations on \textsc{Symbolic Data} are available at \url{http://symbolicdata.org}.


\section{Current Developments in the \textsc{Symbolic Data} PRoject}
\label{sctn:SData}

\subsection{The General Idea}

When \textsc{Symbolic Data} started, the main incentive was to revise a largely unstructured collection of polynomial systems from different areas and thus making them more accessible and in consequence easier to maintain. Additionally, the gained knowledge and experience should allow \textsc{Symbolic Data} to be extended to other computer algebra related data. We will begin by sketching the main problems very briefly and will then focus on the newer development which draws heavily from the rapid development of the Semantic Web technologies in the last years.

Right from the beginning of \textsc{Symbolic Data}, it was clear that the data produced by the computer algebra community is highly inhomogeneous. But rather than ignoring this and trying to fit all available data into a given arbitrary schema, we sought ways to incorporate this inhomogeneity.

There are two main questions which arise in the given scenario. The first one concerns the resources themselves: how is the data stored? Since there is a wide range of computer algebra systems, the file formats vary as well. The problem was here, to find a suitable file format that is versatile enough to serve as a lingua franca between different computer algebra systems and as well widespread enough to be readily processed by most programming languages. This question concerns the resources themselves and will be discussed in subsection \ref{XML_resources}.

The second question is where, how and which information \textit{about} the resources is stored. This kind of information is often called \textbf{metadata}. \textsc{Symbolic Data} stores data and metadata separately. This has two advantages: Firstly, data and metadata do not have to be stored on the same server. Secondly, for resources with large file sizes it is often tedious to move around and edit those resource files.

The question of \textit{how} to store metadata will be treated in subsections \ref{different_ways} and \ref{RDF_metadata}. The hardest part remains \textit{which} metadata should be stored. For polynomial systems there are concepts like dimension, highest degree, number of variables, homogeneity properties etc. For other fields it will be other concepts and this can only be answered by the computer algebra community as a whole.

\subsection{XML as Resource Format}
\label{XML_resources}

Until around the year 2000 the \textsc{XML} file format was very popular to store any kind of data. It still is but there are other alternatives, most prominent \textsc{JSON} and \textsc{YAML}. One key property of all three is that the files are human-readable. From a programmer's point of view, an important feature is the availability of \textsc{XML} parsers and serializers for every major programming language. 

It should be noted that \textsc{XML} is not really a language on its own but rather a set of rules that can be extended by an \textsc{XML} Schema (which usually consists of one or more XSD files) to define a custom language or file format. One of the various applications of this technique is for instance XHTML (\cite{pemberton2000xhtml}). 

Another advantage is that an \textsc{XML} file can be easily checked against an \textsc{XML} Schema to determine whether it is well-formed or not. \textsc{Symbolic Data} uses \textsc{XML} based file formats to store resource data.\footnote{We note that MathML is already an \textsc{XML} based file format for mathematical data. We decided not to use MathML for our resources. MathML describes formulas in a very detailed and almost scrupulous way that creates a big overload in terms of file size and also greatly decreases the human readability.} A recent discussion has been if one should adhere to this or if this question is not even relevant. With the separation of resource data and metadata this is however not a question that has to be answered right away.

\subsection{Different ways to collect metadata}
\label{different_ways}

There are mainly three ways to create and store metadata, each of which with its own advantages and disadvantages. 

A common way to store information about given resources is to set up a `classical' relational database, e.g. MySQL or PostgreSQL. However, before setting up a database the structure of the data should be clear to the point that an Entity-relationship model could be drawn. Altering the abstract model after a large amount of data has already been stored in the database is often a tedious task, to say the least.

Another, somewhat different, method to create metadata is to add tags or keywords to each record. This has been made popular by the Web 2.0. The main advantage is, that additional tags can be introduced at any given time. One disadvantage concerning consistency is that an author of an entry is not forced to include these tags.

While \textsc{XML} was not originally aimed at being a format for databases, it can be used to do precisely this. And just as SQL is a languange to query SQL databases, there is also a query language for \textsc{XML} databases, namely XQuery.

The third way to store metadata is to use the RDF/OWL standards. While SQL databases can be visualized as collection of tables with cross references, RDF metadata can be seen as a directed graph with labelled edges. Since this is the way we adopted for \textsc{Symbolic Data} we will discuss this in greater detail in the next subsection.

\subsection{Collecting meta data with RDF}
\label{RDF_metadata}

Even with the extended notion of tags that \textsc{XML} can establish, there is still something missing. Tags are often not independent but bear a structure themselves. For instance, a given tag can be a more special version of another tag. These relations between tags are often called `semantics'.

RDF is an acronym of ``Resource Description Framework'' and it is above all a data model. There are different ways to serialize RDF, so RDF data can be written as \textsc{XML} or JSON, as well as Turtle (``Terse RDF Triple Language'') which is a format specifically designed for RDF. 

The basic idea is that each piece of information is stored in form of so-called \textbf{triples}. Each triple consists of a subject $s$, a predicate $p$ and an object $o$. Those terms are used rather loosely instead of adhering to strict linguistic rules. 
Written down in Turtle, a triple can be expressed by simple juxtaposition and a final period:

\begin{equation} s \; p \; o \; . \label{spo} \end{equation}

Subjects and predicates have to be a URI (Uniform Resource Identifier (\cite{berners1994universal})) while objects (or `values') can be either be a URI or a literal in lexical form (a text string) which can either be plain or typed. There are some common types (e.g. `integer') but custom types can be defined as well. The sets of subjects, objects and predicates are not necessarily (mutually) disjoint. Most notable, a predicate of one triple can also become the subject or object of another triple. It will shortly be clear how this makes sense.

A valid question is of course how one would start working with RDF. Suppose one has no database at all and just a resource at

\url{http://symbolicdata.org/XMLResources/IntPS/Caprasse.xml}.
\newline
One could jot down some properties that help to find this resource. Some of these properties could be extracted directly from the resource, others would have to be calculated. It could look like this:
\begin{verbatim}
<http://symbolicdata.org/Data/PolynomialSystems/Caprasse>
    a sd:IntPS ;
    sd:hasDegree "56" ;
    sd:hasDegreeList "3,3,4,4" ;
    sd:hasLengthsList "4,4,9,9" ;
    sd:hasVariables "x,y,z,t" ;
    sd:relatedXMLResource
      <http://symbolicdata.org/XMLResources/IntPS/Caprasse.xml> .    
\end{verbatim}
Some explanations: These are six triples which could also be written in the form (\ref{spo}). Since all triples share the same subject (the first line), the code can be compacted by ending lines with a semicolon instead of a period. This signifies that the subject in the next triple is the same and thus can be omitted. The \texttt{sd:} is just an abbreviation for \url{http://symbolicdata.org/Data/Model/}.

One great thing about RDF is that one does not have to worry too much about the used predicates. Refactoring data can easily be done by `search and replace'. In many cases it is not entirely clear from the beginning what the `right' predicates are (i.e. the predicates that are best suited). Contrary to an SQL database, RDF supports and even encourages a bottom-up approach to building a database.

\subsubsection{RDFS and OWL.} While RDF mostly defines how the data is structured on the lowest level, there is also RDF Schema, in short RDFS, which extends RDF by defining a number of basic classes and properties. The most basic extension is probably \texttt{rdfs:label}, which can be used to add a `display name' to the URIs.

RDFS also provides the formal notion of an RDF class \texttt{rdfs:Class}, so it is possible to talk about the RDF classes in RDF itself. Building on this, RDFS defines \texttt{rdfs:subClassOf} as well as \texttt{rdfs:domain} and \texttt{rdfs:range}. The latter two are used to define the domain and range of a predicate $p$, in Turtle it looks like this:
\[ p \quad \texttt{rdfs:domain} \quad c \; \texttt{.} \]
where $c$ is an \texttt{rdfs:class}.

The next step is to use OWL (Web Ontology Language) and OWL2 to define a highly elaborated structure of the classes and predicates used. Again, like with RDFS, OWL is not really a new language but it defines new keywords that extend RDF.\footnote{The reason for the new names is partly a historical one. The names mark different stages of the development of the Semantic Web and refer to different specifications.}

The collection of classes, predicates, their relation to each other, and the rules of their application is called \textbf{Ontology}. There are already various ontologies available\footnote{see also \url{http://semanticweb.org/wiki/Ontology}}, most notably the Dublic Core ontology, which defines how to talk about different kinds of publications, and the FOAF ontology, which defines how to talk about persons, their contact information and their connections to other people. We will not go into further detail about the vast range of possibilities that OWL offers.

One key idea of the Semantic Web is, that one does not necessarily invent new ontologies but use and possibly extend existing ones. But even if one has two similar databases that each use their own ontology, OWL offers predicates to map the vocabulary of one ontology to that of another.

\subsubsection{Putting all into practice with SPARQL.} When working with RDF, the equivalent to a database is a collection of triples that are loaded into a so-called \textbf{triplestore}. When talking about SQL and \textsc{XML} databases we also mentioned their respective query languages. With SPARQL there is also a query language for RDF\footnote{Actually there is more than one query language for RDF, but we focus on SPARQL here. For details see \url{http://en.wikipedia.org/wiki/RDF_query_language}}. We will not explain SPARQL in greater detail here, but many examples can be found and tested live at \url{http://symbolicdata.org/wiki/QuickStart}.

For the moment, let's just note that with SPARQL one can formulate queries like
\begin{itemize}
\item find all resources have a certain property (e.g. a degree of at most 36)
\item find all resources that are missing a certain predicate $p$
\item find all resources that are in class $c_1$, but not in class $c_2$
\item find all resources of class $c_1$ together with their properties defined by predicates $p_1, \ldots, p_n$
\end{itemize}
SPARQL returns a subset of the triples from the triplestore. The format can be JSON, \textsc{XML}, Turtle and some others. What format is suited best depends on how the results will be further processed. 

\subsubsection{Linked Data.} With RDF, OWL and SPARQL it is possible to advance beyond single databases towards a linked data network of computer algebra resources. This is of course a work in progress which can neither be done overnight nor by a single person. RDF is no magic tool which does all this automatically, but what it does is to provide the means to seriously tackle this project.

\subsection{Ways of Using and Contributing to \textsc{Symbolic Data}}

After having outlined the principles and benefits, we will now present the possibilities of using \textsc{Symbolic Data} and contributing to the project. Depending on the background and the time one is able to invest, there are different methods.

One of our goals is to incorporate new metadata into \textsc{Symbolic Data}, but even more we are interested in getting more people involved with it. The first place to start to get more information is our Wiki at
\url{www.symbolicdata.org}.


For people who are already familiar with RDF or want to become acquainted with it, there is our official repository at GitHub:
\url{github.com/symbolicdata/symbolicdata}.
It contains some resource data and of course metadata. One can fork this repository, include own metadata and send a pull request.

Even without generating RDF metadata, it is possible to use the existing metadata and set up a local triplestore with (a subset of) \textsc{Symbolic Data}. Custom SPARQL queries can provide exactly the presentation of the data a user requests.

One might also already have a relational database. In this case, there are tools that can easily convert an SQL database into RDF and merge it with the collection of \textsc{Symbolic Data}. \textsc{Sparqlify}\footnote{\url{http://aksw.org/Projects/Sparqlify.html}} even does this `live' by translating SPARQL queries to SQL queries.

On the other hand, one might also have a large set of resources in a well-defined format for which it is (at least implicitly) clear how to calculate or extract metadata. We can help to do this (semi)automatically and store the result in RDF which can then be converted into an HTML presentation while additionally providing a SPARQL interface to others.



\section{SDEval}
\label{sctn:SDEval}

\subsection{Particularities about Benchmarking in Computer Algebra}

\subsubsection{Challenges.}

Writing benchmarks in the field of computer algebra differs from other
benchmarking tasks. A collection of challenges that appear is the following.
\begin{itemize}
  \item Sometimes, the results of computations are not unique; that
    is, several nonequal outputs can be equivalently correct. It is
    not always possible to find a canonical form for an output. Even
    if this is the case, the transformation of output into the
    canonical form can be quite costly. Moreover, the latter
    transformation is not necessarily provided by every single
    computer algebra system.
  \item Related to the previous item: If an answer is not unique, then
    the evaluation of the correctness of the output is often far from trivial.
  \item The field of computer algebra deals with a large variety of
    topics, even though it can be divided into classes of areas where
    certain common computational problems do appear. Thus, there need
    to be collections of benchmarks, optimally one as a standard for each class. The benchmark creation process should be flexible to be applicable in a wide range of areas.
\end{itemize}

We tried to address those challenges as much as possible when designing our toolkit.

In particular, the first item is something that differs the creation of benchmarks for computer algebra problems from most other fields of studies.

The second item leads to one of the design decisions we made for
\textsc{SDEval}, namely that we provide an interface for decision
routines, and partially include some serving as an example how they
could be added. Then, a particular community can deal with this
question based on their problems, and provide \textsc{SDEval} with
the information on what routine to call to obtain an answer.

\subsubsection{Correct and Feasible Time Measurement.} Another seemingly trivial, yet controversial question is the correct
time measure of computations, as mentioned in the introduction. It is very common in computer algebra
systems to provide a time measuring functionality, and many of the
timings provided in papers were calculated using those commands, since
it is easily available.

Nevertheless, this methodology is questionable. Often one
cannot verify their validity due to e.g. their source not being
open. Furthermore, sometimes some run-time-benefiting calculations are
already done during the initialization phase; therefore one has to
specify clearly where to start the provided time measurement. If one makes use of the implemented techniques,
every program has to be analyzed in detail to find the correct spot to
start the time counting in order to make the
comparison fair. Hence, the use of system-provided time measuring is not practical for fair comparisons. 

A widely-spread method in software
development is to run programs with the \texttt{time} command provided
with \textsc{Unix} based operating systems
. Even though the time for
parsing input -- which is in general not the complex part about the
computations done in computer algebra -- would then also being taken into account, we decided that this
method is the best choice for \textsc{SDEval}.

It has also another
benefit: We are interested in extracting the timing results from the
output files in an automated way, and there is a standard for providing timings given by
the \textsc{IEEE} standard \texttt{IEEE Std 1003.2-1992
  (``POSIX.2'')}); the \texttt{time} command can be instrumented using
a parameter to
provide its output according to this standard. Arranging this format
for the output
with the help of the included time measurement mechanisms in computer
algebra systems can be regarded as an infeasible requirement for a user.

\subsection{The Creation of a Benchmark Suite}
\subsubsection{Basic Terminology.}

Let us start with defining some terminology we want to use throughout this section. This will serve the purpose of a better understanding of the design principles of \textsc{SDEval}.


\begin{definition}[SD-Table]
\label{def:SD-Table}
\index{SD-Table}
An SD-Table denotes the folder structure of a chosen sub-folder in the
\texttt{XMLRessources} folder in the \textsc{Symbolic Data} project. Those sub-folders represent the tables with computation problems in the \textsc{Symbolic Data} project.
\end{definition}

\begin{example}[SD-Table]
An example for an SD-Table is \texttt{IntPS}. It contains instances of ideals in a
     polynomial ring over $\mathbb{Q}$ using integer coefficients.
\end{example}

\begin{definition}[Problem Instance]
\label{def:ProblemInstance}
\index{Problem Instance}
A problem instance is in our context a representation
of a concrete input -- aligned to the \textsc{Symbolic
Data} format -- that can be used for one or more algorithms. The input values for the chosen algorithm are contained in this
problem instance. A problem instance is always contained in an SD-Table.
\end{definition}

\begin{example}[Problem Instance]
  A problem instance is for example the entry \texttt{Amrhein} (an integer polynomial system taken from \cite{amrhein1996walking}) in the
  SD-Table \texttt{IntPS}. It contains variables and a basis of
  polynomials, and those can be used for Gr\"obner basis computations,
  for example.
\end{example}

\begin{definition}[Computation Problem]
  \label{def:ComputationProblem}
  \index{Computation Problem}
  A computation problem is a concrete and completely specified member
  of a family of algorithms.
 In the context of \textsc{SDEval}, it specifies which computations
  we want to perform on certain problem instances.

  A selection of computation problems is already provided in the SD-Table
  \texttt{COMP}. The selection can be extended by the user.
\end{definition}

\begin{example}[Computation Problem]
 A computation
problem is for example the computation of a Gr\"obner basis given an
ideal over a polynomial ring over $\mathbb{Q}$ using the
lexicographic ordering (abbr. \texttt{GB\_Z\_lp}, can be found in the
SD-Table \texttt{COMP}).
\end{example}

\begin{definition}[Task]
  \label{def:Task}
  \index{Task}
  A task consists of a computation problem and a selection of problem instances
  that are suitable as inputs for it.
\end{definition}

\subsubsection{Automated Creation of Benchmarks}

Now that we have defined some basic terminology, we will address how a
benchmark suite can be generated using the problem instances given in
the SD-Tables.

In the toolbox, one can find two \textsc{Python} programs that can do this job: \texttt{ctc.py}
  and \texttt{create\_tasks\_gui.py}.

The first one is a command-line program, the
second one provides a graphical user interface.

\begin{figure}
\caption{The selection of the problem instance from integer polynomial
systems}
\label{fgr:selectPI}
\begin{center}
\includegraphics[width=.8\textwidth]{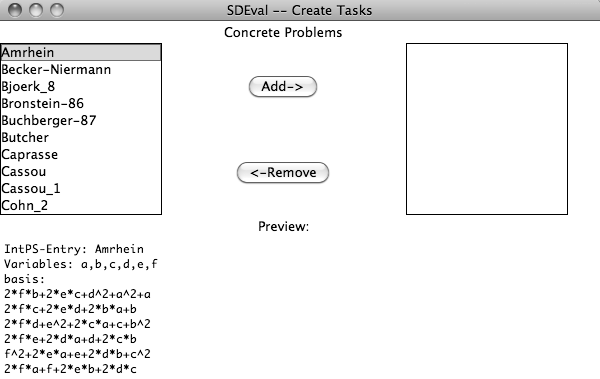}
\end{center}
\end{figure}

Those scripts perform the following three steps
\begin{enumerate}
  \item The user chooses from a set of computation problems.
  \item After that, the script collects possible problem instances
    across the SD-Tables and presents them to the user. One can pick
    the desired problem instances that should be included in the
    benchmark. An illustration of this step is given in Figure
    \ref{fgr:selectPI}.
  \item In the last step, besides setting configuration parameters, the user selects from a set of computer
    algebra systems for which it is known that they contain
    implementations of the algorithms that solve the selected
    computation problem.
\end{enumerate}

After these three steps, the user confirms his choices and
a folder is generated containing executable codes for the selected
computer algebra systems, a \textsc{Python} script to run all the calculations
and some adjustable configuration files (e.g. if the user wants to
change call parameters for a computer algebra system). The concrete structure is given as in Figure \ref{fgr:TaskFolder}.

\begin{figure}
\caption{Folder structure of a taskfolder}
\label{fgr:TaskFolder}
{\scriptsize{
\begin{verbatim}
+ TaskFolder
| - runTasks.py         //For Running the task
| - taskInfo.xml        //Saving the Task in XML-Structure
| - machinesettings.xml //The Machine Settings in XML form
| + classes             //All classes of the SDEval project
| + casSources          //Folder containing all executable files
| | + SomeProblemInstance1
| | | + ComputerAlgebraSystem1
| | | | - executablefile.sdc
| | | + ComputerAlgebraSystem2
| | | | - executablefile.sdc
| | | + ...
| | + SomeProblemInstance2
| | | + ...
| | + ...
\end{verbatim}}}
\end{figure}

We will refer to this folder as taskfolder from now on. This folder
can then be sent to the machine where the computations are intended to
be run.

As outlined before, the creation tool is very flexible and easily
extensible. This is due to the object oriented nature of the code
written in \textsc{Python}. One can specify new computation problems,
and declare which problem instances can be chosen as inputs. The
respective code for the computer algebra systems can be added in a
template-fashion and does not require familiarity with the particular concepts of \textsc{Python}.

\subsection{Running a Benchmark Suite}

\textbf{General Assumption 1:} Whereas the creation of the benchmark
suite is possible on any machine where \textsc{Python} is installed,
the running routine requires a machine running with a
\textsc{UNIX}-like operating system (e.g. \textsc{Linux} or
\textsc{Mac OS X}). We require the \texttt{time}
command or some equivalent to be supported, which is in general always the case on
\textsc{UNIX} systems.

\noindent\textbf{General Assumption 2:} Calculations are run within a
terminal. This decision was made due to the fact that calculations are often sent to a compute server. The
connection to that server is in general provided through a terminal interface.
\vspace{-6pt}

The running of a benchmark is closely connected to the taskfolder as
presented in the previous section. As one can see in Figure
\ref{fgr:TaskFolder}, it contains a \textsc{Python} script called
\texttt{runTasks.py}. One can either generate an individual taskfolder
using the design principles given in the documentation, or one can use
a taskfolder generated by the task creation scripts.

If one executes \texttt{runTasks.py}, all the stored scripts for all the contained computer algebra systems will be run consequently. Using execution parameters, one can instruct the script to kill a process once a given time or memory limit is reached.

The script will create -- if not yet existent -- a sub-folder within
the taskfolder named \texttt{results}. Within \texttt{results}, there
will be a folder named by the time stamp when \texttt{runTasks.py} was
executed, where it will store the results of the computations and some
monitoring information about the executed scripts in form of \textsc{HTML} and \textsc{XML} files.

During the execution process, the user can feel free to terminate manually a running process without having to restart \texttt{runTasks.py}. It will simply continue with the next waiting program on the next script in the queue.

This design of the benchmark execution part has the following
benefit. Future authors that execute their scripts on certain files
could provide their taskfolder with the paper they submitted. Then
everyone can see the results (i.e. the outputs of the programs), and
verify the timings using the calculated table. Furthermore, they can
run the calculation using \texttt{runTasks.py} after adjusting the
configuration to their machine (i.e. replacing the call commands for
the computer algebra systems to those used on one's machine).


There are further uses of the running routines. As one can see, the execution of the benchmarks can be seen completely
detached from the creation part. This means, that one can create one's
own taskfolder, defining programs one wants to run and provide the
inputs inside the \texttt{casSources} folder.

Even though the routines were designed to fit especially the needs of the
computer algebra community, the principles can be used for almost any
kind of program.

Another use would be to keep track of the development process of a
software project over time. Executing the \texttt{runTasks.py} script after
every version change would reveal profiling information on the
different examples. The profiling can be automatized since the
timing-data after every run is stored in an \textsc{XML} file.

\subsection{Ways of Customizing and Contributing to \textsc{SDEval}}

We have seen in the last section that the part of the execution of the
respective programs on the problem instances is highly
customizable. There are also ways for customization of the part where one creates benchmarks.

For the case that we have not considered a certain computer algebra system that is capable of solving a particular computation problem, a user familiar with this computer algebra system is able to provide a template without the need of a deep knowledge of \textsc{Python}.

For the case that there exists a not yet considered computation problem for which inputs can be derived from existing SD-tables, one can provide a representative \textsc{Python} class, a template for a computer algebra system code and link the respective SD-Tables to it.

More possibilities and details can be found in the documentation of \textsc{SDEval}.

\section{Related Work}
\label{sctn:relatedWork}

\subsection{Related Work to \textsc{SDEval}}
\textsc{StarExec} (\cite{stump2012introducing}): This is an infrastructure especially for the logic solver communities. Its main focus is to provide a platform for them to manage their benchmark libraries and run solver competitions. It is widely used in conferences based on logic solving to evaluate the benefits of new approaches. Moreover, it includes translators of problems between the different communities dealing with logic solving.

\textsc{homalg} (\cite{barakat2008homalg}): Focusing on constructive homological algebra, the \textsc{homalg} project provides an abstract structure for abelian categories and is distributed as a package of the computer algebra system \textsc{GAP} (\cite{GAP4}). For time critical computations, it allows the usage of other computer algebra systems, i.e. the task is translated to the respective system and then executed. 

\textsc{Sage} (\cite{stein2008sage}): The popular computer algebra system \textsc{Sage} provides as an optional package an interface to the database of integer polynomial systems (\texttt{IntPS}) of the \textsc{Symbolic Data} project. One can directly load those problem instances as objects in \textsc{Sage} for further calculations. 

\subsection{Related Work to \textsc{Symbolic Data}}

There are various sets of problem instances for different computation problems collected by different communities during the last couple of decades. Here is a selection of some interesting projects.

\textsc{PoCab} (\cite{samal2012pocab}): \textsc{PoCab} is a collection of models coming from the field of biology and chemistry. They concentrate on examining the different algebraic entities given in those models in order to apply algebraic methods in the process of their analysis. Their data of interest is coming from two renowned and publicly available databases, namely \textsc{KEGG} (\cite{kanehisa2000kegg}) and the \textsc{Biomodels Database} (\cite{le2006biomodels}).

\textsc{Polytope Database} (\url{www.mathematik.tu-darmstadt.de/~paffenholz/data.html}): 
We established an interesting contact, namely to Dr. Andreas Paffenholz. He has a large amount of resource data as Polymake files for different kinds of polytopes. We will extract and possibly generate metadata and include this into \textsc{Symbolic Data}.

\textsc{swMATH} (\url{http://swmath.org}): The focus of this project is to serve as an information service for mathematical software. It consists of an extensive database of software projects from the field of mathematics, and contributes a systematic linking of program packages with relevant publications.

\textsc{Qaos} (\url{http://qaos.math.tu-berlin.de}): This project at the TU Berlin is the successor of the ``Kant Database of Number Fields''. It provides an interface to various categories of algebraic objects. 


\section{Conclusion and Future Work}
\label{sctn:Conclusion}

We have presented the latest developments of the \textsc{Symbolic Data} project, and a benchmarking tool named \textsc{SDEval} that is built on top of it. In this paper, we addressed the particularities of benchmarking in the field of computer algebra, and with \textsc{SDEval}, we have presented a flexible, extensible and easy-to-use tool that is designed to accept the challenge.

Moreover, we introduced a practice how the reproduction and the analysis of computations with their timings would become more feasible in the future.  Our approach for that is the taskfolder containing the benchmark program and the respective input files.

A future task will be to extend the benchmark creation tool to contain both more computer algebra systems and computation problems. The program that runs the benchmarks will include output interpretation routines to determine reliably the correctness of results. For that, one has to consider every computation problem in a detailed way and we hope for support from the communities in the future to accomplish that.

The future work on the \textsc{Symbolic Data} project will consist of expanding the database with new entries provided by the community. We are constantly establishing contacts to interested researchers.
Prof. Dr. J\"urgen Kl\"uners for example, who already has a huge Postgres database of number fields\footnote{\texttt{http://galoisdb.math.uni-paderborn.de/home}}, pointed us to other resources for local fields and number fields, which are not yet conveniently accessible.

Prof. Dr. Max Horn described briefly to us how a \textsc{Symbolic Data} library could be included into \textsc{Gap}. Since there are also resources available through \textsc{Gap} databases, it would be also interesting also make these searchable with \textsc{Symbolic Data}. 

In personal talks with Pavel Metelitsyn it became clearer that a translation of resource data is an important task. Also it is interesting to not only store metadata but to generate it when needed. \textsc{Sage} (\cite{stein2008sage}) is probably best suited as an additional software service that can be combined with \textsc{Symbolic Data}. We will investigate this idea further.

As benchmarking is a very wide-ranged topic, we will figure out in the future if there are more challenges -- maybe caused by computation problems we have not considered yet -- that we are  not aware of in the present state. It remains a practically relevant and interesting problem.

\section*{Acknowledgements}
We thank the ``Deutsche Forschungsgesellschaft'' (DFG) for partial financial support for the development of \textsc{SDEval} (``DFG Priority Project SPP 1489''). 

We thank the ``Europ\"aische Sozialfonds f\"ur Deutschland'' (ESF) for
funding the work on \textsc{Symbolic Data} in the context of the
project ``eScience – Forschungs\-netz\-werk Sachsen''. Moreover we
personally thank Dr. Toni Tontchev for his enthusiastic support.

We are grateful to Hans-Gert Gr\"abe for his encouragement and the
fruitful discussions we had with him.

The authors thank Johannes Waldmann for his presentation on the
benchmarking practice of the logic solver communities during the
\textsc{Symbolic Data} workshop.

Special thanks to Mark Giesbrecht for his helpful suggestions and comments.



\bibliography{cascPaper}

\newcommand{\etalchar}[1]{$^{#1}$}
\begin{thebibliography}{LNBB{\etalchar{+}}06}

\bibitem[AGK96]{amrhein1996walking}
Beatrice Amrhein, Oliver Gloor, and Wolfgang K{\"u}chlin.
\newblock Walking faster.
\newblock In {\em Design and Implementation of Symbolic Computation Systems},
  pages 150--161. Springer, 1996.

\bibitem[BCD{\etalchar{+}}11]{barrett2011cvc4}
Clark Barrett, Christopher~L. Conway, Morgan Deters, Liana Hadarean, Dejan
  Jovanovi{\'c}, Tim King, Andrew Reynolds, and Cesare Tinelli.
\newblock Cvc4.
\newblock In {\em Computer Aided Verification}, pages 171--177. Springer, 2011.

\bibitem[BH08]{bjorck2008all}
Goran Bjorck and Uffe Haagerup.
\newblock All cyclic p-roots of index 3, found by symmetry-preserving
  calculations.
\newblock {\em arXiv preprint arXiv:0803.2506}, 2008.

\bibitem[BL94]{berners1994universal}
Tim Berners-Lee.
\newblock Universal resource identifiers in www.
\newblock 1994.

\bibitem[BR08]{barakat2008homalg}
Mohamed Barakat and Daniel Robertz.
\newblock {\sc homalg} --a meta-package for homological algebra.
\newblock {\em Journal of Algebra and its Applications}, 7(03):299--317, 2008.

\bibitem[BST10]{barrett2010smt}
Clark Barrett, Aaron Stump, and Cesare Tinelli.
\newblock The smt-lib standard: Version 2.0.
\newblock In {\em Proceedings of the 8th International Workshop on
  Satisfiability Modulo Theories (Edinburgh, England)}, volume~13, 2010.

\bibitem[DMB08]{de2008z3}
Leonardo De~Moura and Nikolaj Bj{\o}rner.
\newblock Z3: An efficient {SMT} solver.
\newblock In {\em Tools and Algorithms for the Construction and Analysis of
  Systems}, pages 337--340. Springer, 2008.

\bibitem[GAP13]{GAP4}
The GAP~Group.
\newblock {\em {GAP -- Groups, Algorithms, and Programming, Version 4.6.3}},
  2013.

\bibitem[Gr{\"a}09]{grabe2006symbolicdata}
Hans-Gert Gr{\"a}be.
\newblock The {\sc symbolicdata} project.
\newblock Technical report, Technical report (2000-2009), 2009.

\bibitem[KFI{\etalchar{+}}87]{katsura1987distribution}
S.~Katsura, W.~Fukuda, S.~Inawashiro, N.~M. Fujiki, and R.~Gebauer.
\newblock Distribution of effective field in the ising spin glass of the $\pm
  j$ model at {$T= 0$}.
\newblock {\em Cell Biophysics}, 11(1):309--319, 1987.

\bibitem[KG00]{kanehisa2000kegg}
Minoru Kanehisa and Susumu Goto.
\newblock {KEGG}: {K}yoto encyclopedia of genes and genomes.
\newblock {\em Nucleic acids research}, 28(1):27--30, 2000.

\bibitem[LNBB{\etalchar{+}}06]{le2006biomodels}
Nicolas Le~Novere, Benjamin Bornstein, Alexander Broicher, Melanie Courtot,
  Marco Donizelli, Harish Dharuri, Lu~Li, Herbert Sauro, Maria Schilstra, Bruce
  Shapiro, et~al.
\newblock {BioModels} {D}atabase: a free, centralized database of curated,
  published, quantitative kinetic models of biochemical and cellular systems.
\newblock {\em Nucleic acids research}, 34(suppl 1):D689--D691, 2006.

\bibitem[Neu12]{sneumann2012Parallel}
Severin Neumann.
\newblock Parallel reduction of matrices in gr{\"o}bner bases computations.
\newblock In Vladimir~P. Gerdt, Wolfram Koepf, Ernst~W. Mayr, and Evgenii~V.
  Vorozhtsov, editors, {\em Computer Algebra in Scientific Computing}, volume
  7442 of {\em Lecture Notes in Computer Science}, pages 260--270. Springer
  Berlin Heidelberg, 2012.

\bibitem[P{\etalchar{+}}00]{pemberton2000xhtml}
Steven Pemberton et~al.
\newblock Xhtml™ 1.0 the extensible hypertext markup language.
\newblock {\em W3C Recommendations}, pages 1--11, 2000.

\bibitem[S{\etalchar{+}}08]{stein2008sage}
William Stein et~al.
\newblock {\sc Sage}: Open source mathematical software, 2008.

\bibitem[SEW12]{samal2012pocab}
Satya~Swarup Samal, Hassan Errami, and Andreas Weber.
\newblock {PoCaB}: a software infrastructure to explore algebraic methods for
  bio-chemical reaction networks.
\newblock In {\em Computer Algebra in Scientific Computing}, pages 294--307.
  Springer, 2012.

\bibitem[SST12]{stump2012introducing}
Aaron Stump, Geoff Sutcliffe, and Cesare Tinelli.
\newblock Introducing starexec: a cross-community infrastructure for logic
  solving.
\newblock {\em Comparative Empirical Evaluation of Reasoning Systems}, page~2,
  2012.

\end{thebibliography}

\end{document}